\documentclass[%
preprint,
superscriptaddress,
frontmatterverbose, 
amsmath,amssymb,
]{revtex4-2}

\usepackage{graphicx}
\usepackage{dcolumn}
\usepackage{bm}
\usepackage{graphicx} 
\usepackage{xcolor} 
\usepackage{hyperref}
\hypersetup{
    colorlinks=true, 
    linktoc=all,     
    linkcolor=blue,  
    citecolor=blue,
    filecolor=blue,
    urlcolor=blue
}
\graphicspath{ {./} }
\usepackage{wrapfig}
\usepackage{amsmath}
\usepackage{comment}
\usepackage{amssymb}
\usepackage{amsmath}
\usepackage [english]{babel}
\usepackage [autostyle, english = american]{csquotes}

\usepackage[normalem]{ulem}

\usepackage{ragged2e}
\newcommand{\kbf}{{\mathbf{k}}}
\usepackage[font=small,
   justification=RaggedRight,
   format=plain]{caption}
\usepackage{lineno}
\usepackage{amsmath}  
\usepackage{etoolbox} 

\newcommand*\linenomathpatch[1]{%
  \cspreto{#1}{\linenomath}%
  \cspreto{#1*}{\linenomath}%
  \csappto{end#1}{\endlinenomath}%
  \csappto{end#1*}{\endlinenomath}%
}

\linenomathpatch{equation}
\linenomathpatch{gather}
\linenomathpatch{multline}
\linenomathpatch{align}
\linenomathpatch{alignat}
\linenomathpatch{flalign}

\begin{document}

\preprint{APS/123-QED}

\title{Oscillations and confluence in three-magnon scattering of ferromagnetic resonance}

\author{Tao Qu}
     \affiliation{
     Department of Electrical and Computer Engineering, University of Minnesota, Minneapolis, MN, 55455, USA \\
    }
\author{Alex Hamill}%
    \email{hamil483@umn.edu}
    \affiliation{%
     School of Physics and Astronomy, University of Minnesota, Minneapolis, MN, 55455, USA \\
    }%
\author{R. H. Victora}
     \affiliation{
     Department of Electrical and Computer Engineering, University of Minnesota, Minneapolis, MN, 55455, USA \\
    }
\author{P. A. Crowell}%
\affiliation{%
 School of Physics and Astronomy, University of Minnesota, Minneapolis, MN, 55455, USA \\
}%

\begin{abstract}


We have performed a time-resolved and phase-sensitive investigation of three-magnon scattering of ferromagnetic resonance (FMR) over several orders of magnitude in excitation power. We observe a regime that hosts transient oscillations of the FMR magnon population, despite higher-order magnon interactions at large powers. Also at high powers, the scattering generates $180^\circ$ phase shifts of the FMR magnons. These phase shifts correspond to reversals in the three-magnon scattering direction, between splitting and confluence. These scattering reversals are most directly observed after removing the microwave excitation, generating coherent oscillations of the FMR magnon population much larger than its steady-state value during the excitation. Our model is in strong agreement with these findings. These findings reveal the transient behavior of this three-magnon scattering process, and the nontrivial interplay between three-magnon scattering and the magnons' phases.


\end{abstract}

\maketitle


Magnons are the quanta of collective spin excitations. Their phase degree of freedom, highly nonlinear behavior \cite{rezende1990spin,wigen1990route,laulicht1999transient,mathieu2003brillouin,an2004high,kabos1994measurement,krawiecki1995off,carroll1990chaos,carroll1989chaos,carroll1987chaotic,araujo2003dual,wigen1990route,srinivasan1988observation,zakharov1975spin,cherepanov1993collective,rezende1990spin2,slavin1994instability,synogach2000ultrashort,carroll1987chaotic,carroll1989chaos,carroll1990chaos}, and long lifetimes make them an active subject in fundamental research \cite{Verba2021TheoryOT,barsukov2019giant,schultheiss2019excitation,zhou2021magnon} and research towards next-generation microwave and information technology \cite{serga2010yig,adam1993frequency,adam2013magnetostatic,pirro2021advances,barman20212021,csaba2017perspectives,chumak2019fundamentals,etesamirad2021control}. A magnon mode's population (hereafter referred to as its amplitude) can be excited above a threshold value such that it becomes unstable, returning to the threshold value through three-magnon splitting \cite{suhl1957theory}. This nonlinear process is referred to as the first-order Suhl instability.
	
The zero-wavevector magnon mode corresponds to ferromagnetic resonance (FMR) and has a dramatically low threshold amplitude for this instability, particularly in magnetic insulators due to their low damping. This allows for efficient excitation of finite-wavevector magnons and the study of nonlinear magnon interactions over a wide power range. However, little is known about how this instability evolves in time and the role of the excitation power. Previous experiments have observed no associated power-dependence \cite{cunha2015nonlinear}, or have instead focused on the influence of dipole radiation \cite{desormiere1969transient} or of group velocity and proximity to the excitation antenna \cite{liu2019time}. The reverse process of splitting is referred to as confluence, which also requires further investigation \cite{liu2019time}. Its relationship with splitting is an unresolved question, as is the relationship between three-magnon scattering processes and the magnons' phases.

To this end, we employ time-resolved homodyning spectroscopy to examine this instability with phase-sensitivity over five orders of magnitude in microwave excitation power. We observe a regime hosting power-dependent transient oscillations of the FMR amplitude, that at high powers the instability induces $180^\circ$ phase shifts of FMR, and that these phase shifts correspond to reversals between three-magnon splitting and confluence. Furthermore, turning off the microwave excitation stimulates such reversals, generating prolonged and coherent oscillations of the FMR amplitude. Our model is in strong agreement with these observations, and explains the origin of the oscillatory regime as well as the oscillations after turn-off. The oscillatory regime persists up to the highest powers employed, being remarkably robust against the higher-order interactions that arise.

The FMR mode $b_0$ with a frequency $f_0$ is subject to the first-order Suhl instability when magnon modes $b_{\pm \kbf_i}$ at $f_0/2$ are available. This occurs at low FMR frequencies for in-plane magnetized films, due to a minimum in the magnon dispersion \cite{kalinikos1986theory_maintext,mansuripur1988demagnetizing}; an increase in wavenumber suppresses the dynamic demagnetization field. With these modes available, $b_0$ becomes unstable above a threshold amplitude and undergoes three-magnon splitting to $b_{\pm \kbf_i}$. In the reverse process, confluence, two magnons $b_{\pm \kbf_i}$ combine into a magnon $b_0$ [Fig. 1(a)].


\begin{figure}[b]
    \includegraphics{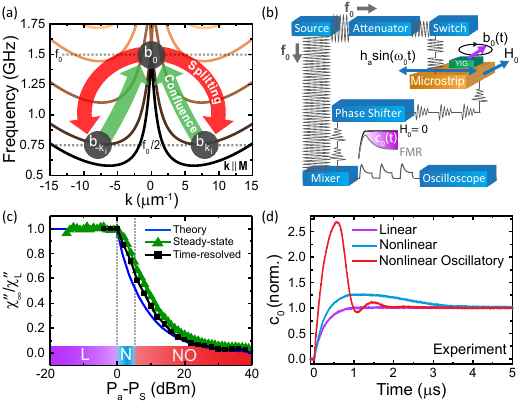}
    \caption{ FIG. 1. (a) The magnon dispersion of $3 \, \mathrm{\mu m}$ YIG films magnetized in-plane, in $15^\circ$ increments of the angle between the wavevector $\kbf$ and the static magnetization $\mathbf{M}$. (b) Schematic diagram of the experiment. (c) $\chi''_\infty(P_a)/\chi''_L$ predicted from Suhl's theory (blue) and that measured through lock-in techniques (green) and time-resolved measurements (black). The color-coded regions correspond to the Linear (L), Nonlinear (N), and Nonlinear Oscillatory (NO) regimes. (d) Overview of the transient behavior of the FMR magnon population $c_0$ in each regime, normalized to the values of $c_0$ at steady state.}
    \label{fig:Fig1}
\end{figure}


We investigated this instability in the time domain through homodyning spectroscopy [Fig. 1(b)]. The microwave excitation at the desired FMR frequency $f_0$ is converted to the desired applied microwave power $P_a$ and to $8 \, \mathrm{\mu s}$ pulses using an attenuator and switch. These pulses enter a wide microstrip waveguide, generating a spatially uniform microwave magnetic field of amplitude $h_{a}$ throughout the sample: a $3\, \mathrm{\mu m}$-thick film of Yttrium Iron Garnet (YIG). We resonantly excite the sample's FMR mode by matching its FMR frequency with $f_0$, via tuning the static magnetic field $H$ to $H_0$. We primarily investigated the transient behavior at $f_0=1.5$ GHz, with a corresponding resonant field of $H_0=135.1$ Oe. The sample is inductively coupled to the microstrip, such that its FMR response $b_0$ induces a corresponding voltage in the microstrip \cite{maksymov2015broadband}. We obtain the envelope of the microstrip's output voltage by mixing it with a phase- and frequency-matched reference. By subtracting the output voltage's envelope at resonance from that at zero field, we isolate the envelope of the voltage induced by FMR. When nonlinearity-induced phase shifts are absent, this envelope directly corresponds to the FMR amplitude $c_0(t)$ with the susceptibility $\chi''(t)=c_0(t)/h_{a}\propto c_0(t)/\sqrt{P_a}$. For additional experimental details, see Sec. \ref{sec:exp_supp} of the supplementary material (SM).

To verify our experiment, we compare the measured steady-state susceptibility $\chi''_\infty$ with Suhl's theory \cite{suhl1957theory} around the instability's threshold power $P_S$ [see Fig. 1(c)]. For $P_a<P_S$, $c_0$ is in the linear regime, such that its steady-state value is proportional to $h_a$. For $P_a=P_S$, this value corresponds to $c_S$, the threshold amplitude for the (nonlinear) instability regime. The steady-state value of $c_0$ saturates at $c_S$ in the nonlinear regime, as three-magnon splitting occurs for $c_0>c_S$. As such, as $P_a$ is increased, $\chi''_\infty$ decreases as $\chi''_\infty \propto 1/\sqrt{P_{a}}$. We compare this with the experimental results by normalizing $\chi''_\infty$ to its value in the linear regime $\chi''_L$. The experimentally obtained saturation from both steady-state (using lock-in techniques) and time-resolved measurements is in reasonable agreement with the theoretical prediction.

In our time-resolved measurements [Fig. 1(d)], we find both the linear and nonlinear regimes as well as their expected transient behavior: in the linear regime (purple curve) $c_0$ rises monotonically to its steady-state value, while in the nonlinear regime (blue curve) it becomes unstable and then relaxes to $c_S$ via three-magnon splitting. However, we observe an additional regime with a threshold power $P_{osc}>P_S$, in which $c_0$ oscillates at a power-dependent frequency as it relaxes to $c_{S}$ (red curve).

To understand this regime, we have developed a model for three-magnon scattering of FMR in thin films, for the case of resonant excitation by a perpendicular microwave field (see Sec. \ref{sec:model_supp} of SM for details). We derive the associated equation of motion for the circularly-polarized magnetization $m^+$ via the Landau-Lifshitz equation. We then perform a plane-wave expansion of $m^+$ \cite{suhl1957theory} to obtain the equations of motion of the circularly-precessing magnon modes. Afterwards, we employ a classical Bogoliubov transformation \cite{dobin2003intrinsic_main} to obtain the equations of motion of the eigenmodes, the elliptically-precessing magnon modes $b(\kbf)$. We only retain terms up to second-order in the magnon modes, to account for three-magnon scattering while neglecting higher-order interactions. We account for linear damping of $b(\kbf)$ through the relaxation rate $\eta(\kbf)=\omega(\kbf) \epsilon(\kbf)\alpha$ \cite{kambersky1975spin}; $\omega(\kbf)$ is the mode's angular frequency, $\epsilon(\kbf)=\frac{1}{\gamma}\left(\partial \omega / \partial H\right)|_{\kbf,H_0}$ is the ellipticity factor, $\gamma$ is the gyromagnetic ratio, and $\alpha$ is the measured Gilbert damping constant. We only consider the resulting equations of motion of the FMR mode $b_0$ and the $N_k$ magnon modes $b_{\kbf_i}$ of frequency $f_0/2$ in the magnon dispersion. Each of these half-frequency modes' equation of motion is distinguished by $\zeta_{\kbf_i}$, their coupling strength with $b_0$, and their relaxation rate $\eta_{\kbf_i}$. However, these distinguishing parameters have weak variation among the $f_0/2$ modes. Hence, we set $\zeta_{\kbf_i}$ and $\eta_{\kbf_i}$ to their average value over all $f_0/2$ modes, $\bar{\zeta}, \bar{\eta}_k$. This causes the equations of motion for each of the half-frequency modes to be identical, reducing the $N_k$ half-frequency modes to a single effective mode $b_k$. Note that we also equate $b_k(t)$ and $b_{-k}(t)$, as the splitting and confluence processes affect each mode equally. This reduces our model to the two equations of motion
\begin{align}
\dot{b}_0&=(i\omega_0-\eta_0) b_0-N_k \bar{\zeta} b^2_k+\nu h_{a}e^{i(\omega_{0}t-\pi/2)} , \label{eq:db0_dt} \\
\dot{b}_k&=(i\omega_k-\bar{\eta}_k) b_k + \bar{\zeta} b^*_k b_0. \label{eq:dbk_dt}
\end{align}
$\nu$ is the coupling of $b_0$ to the microwave field. The magnon mode's response $b_i$ ($i=0,k$) encodes both its amplitude $|b_i|$ and its phase, such that $b_i(t)=|b_i(t)| e^{i(\omega_i t+\theta_i+\phi_i(t))}$. $\theta_{0,k}=-\pi/2,-\pi/4$ are the modes' phase offsets for the cases of linearity and weak nonlinearity, as found from numerically solving Eqs. \eqref{eq:db0_dt},\eqref{eq:dbk_dt}. $\phi_i(t)$, discussed later, will correspond to phase shifts induced by strong nonlinearity. We define $c_i(t)$ as the mode's amplitude for $\phi_{0,k}=0$, such that $b_i(t)=c_i(t) e^{i(\omega_i t+\theta_i)}$. Inserting this relation into Eqs. \eqref{eq:db0_dt},\eqref{eq:dbk_dt} yields the equations of motion for the magnon mode amplitudes:
\begin{align}
\dot{c}_0&=-\eta_0 c_0-N_k \bar{\zeta} c^2_k+\nu h_{a}, \label{eq:dc0_dt} \\
\dot{c}_k&=-\bar{\eta}_k c_k + \bar{\zeta} c_k c_0. \label{eq:dck_dt}
\end{align}
This definition of $c_0$, as in the experiment, corresponds to the envelope of $b_0$ obtained by mixing it with a frequency-matched reference signal $e^{-i(\omega_0 t+\theta_0)}$. As such, the experiment can be directly compared with the numerical solutions of Eq. \eqref{eq:dc0_dt}. For numerically solving our equations of motion, we set the value of $h_{a}$ such that $h_a/h_{S}$ matches the experiment and simulations; $h_{S}$ is the observed threshold value of $h_a$ for the instability. $\eta_i$ is calculated as described previously. All other parameters are set to the values calculated from our model. See Sec. \ref{sec:model_supp} of SM for their derivations. The initial values are the thermal amplitudes corresponding to the Bose-Einstein distribution (see Sec. \ref{sec:therm_supp} of SM). For details on the simulations and numerical solutions, see Sec. \ref{sec:methods_supp} of SM.


\begin{figure}[t]
    \includegraphics{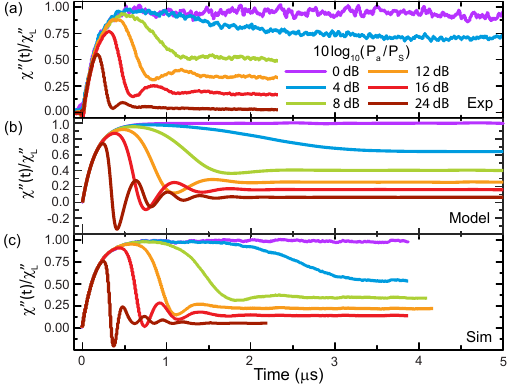}
    \caption{\selectfont{FIG. 2. Comparison of the FMR amplitude's transient behavior from (a) the experiment, (b) the numerical solutions to Eqs. \eqref{eq:dc0_dt},\eqref{eq:dck_dt}, and (c) the micromagnetic simulations. The FMR amplitude $c_0$ is normalized to $\chi''(t)/\chi''_L$ to enable direct comparison between each approach.}}
    \label{fig:Fig2}
\end{figure}


Figure 2 compares the time-evolution of $c_0$ from the experiment, the numerical solutions of Eqs. \eqref{eq:dc0_dt},\eqref{eq:dck_dt}, and the micromagnetic simulations. We normalize to $\chi''(t)/\chi''_{L}$ for direct comparison between each approach. Each curve color corresponds to the same relative power $P_{a}/P_{S}$. The purple and blue curves correspond to the linear and nonlinear regimes, while the green curves correspond to the entrance into the nonlinear oscillatory regime. As $P_a$ is increased, the oscillation frequency $f_{osc}$ monotonically increases while the timescale of the initial transient peak monotonically decreases. Strong qualitative agreement is observed between each approach, which is also the case at 2.5 GHz (see Sec. \ref{sec:2_5GHz_Supp} of SM). The oscillations weaken as one goes from the model to simulation to experiment, presumably due to increasing magnon dephasing. In addition to simulations allowing for other magnon interactions, they include thermal fluctuations which can lead to dephasing. In the experiment, additional dephasing may arise from sample and magnetic field inhomogeneity. 

To analyze the oscillatory regime, we linearize Eqs. \eqref{eq:dc0_dt},\eqref{eq:dck_dt} by Taylor expanding $\dot{c}_0,\dot{c}_k$ about the nonlinear regime's fixed point, which corresponds to steady state; this allows us to treat the second-order terms as negligible. We then impose a time-dependence of the form $c_0,c_k \propto e^{\lambda t}$, and solve for $\lambda$. The transition to the oscillatory regime corresponds to the nonlinear regime's fixed point changing from a stable node to a stable spiral, such that $\lambda$ becomes complex. This transition can be thought of as the point where the splitting rate becomes large enough to produce negative feedback by suppressing $c_0$ to below $c_S$, where splitting is suppressed. This generates exchanges in dominance between the splitting and the microwave excitation terms in Eq. \eqref{eq:dc0_dt}, hence the oscillations. This analysis (see Sec. \ref{sec:linearize_supp} of SM) yields predicted values for the oscillation frequency $f_{osc}$ and the threshold value of $h_a$ for the oscillatory regime, $h_{osc}$:
\begin{align}
f_{osc}=\eta_0 \sqrt{\frac{\Bar{\zeta} \nu}{8\pi^2}(h_{a}-h_{osc})}, \label{eq:fosc_def}\\
h_{osc}=h_{S}(1+\frac{\eta_0}{8\Bar{\eta}_k}). \label{eq:hosc_def}
\end{align}
We compare the predicted scaling from Eq. \eqref{eq:fosc_def} with our results by extracting, via a  Fourier transform, the oscillations' frequency spectra from each approach [Figs. 3(a,b)]. We define $f_{osc}$ at each power as the characteristic peak in the oscillations' frequency spectra and $h_{osc}$ as the value of $h_a$ just below where low-frequency structure is observed in the spectra. For more details, see Sec. \ref{sec:osc_analysis_supp} of SM. The linearized model's predicted scaling $f_{osc}\propto \Tilde{h}^{0.5}$, where $\Tilde{h}=(h_{a}-h_{osc} )/h_{osc}$, is compared to the scaling obtained from each approach [Fig. 3(c)]. We normalize $h_a-h_{osc}$ by $h_{osc}$ to directly compare each approach. The model's results are from the numerical solutions of Eqs. \eqref{eq:dc0_dt}, \eqref{eq:dck_dt}. The filled symbols are those included in the scaling fit (dashed lines) such that quantitative agreement with Eq. \eqref{eq:fosc_def} is observed. The range of agreement for each approach is several orders of magnitude in $P_a$. This is also the case for $f_0=2.5$ GHz (see Sec. \ref{sec:2_5GHz_Supp} of SM). The oscillation frequencies in the experiment and simulations show good agreement, but they are larger than those from our model. This is likely due to an incomplete treatment of damping and/or an underestimation of $\Bar{\zeta}$, as we neglect the full Gilbert damping term and spatial variation of the longitudinal magnetization component.


\begin{figure}[b]
    \includegraphics{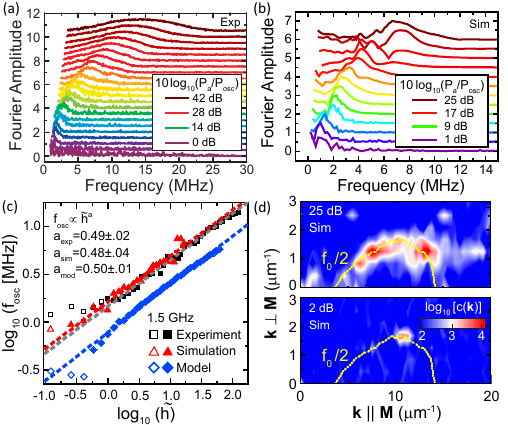}
    \caption{ \selectfont{FIG. 3. Oscillation analysis at $f_0=1.5$ GHz. (a,b) Normalized and offset frequency spectra of the oscillations in (a) the experiment, and (b) the simulations, with a 2 dB increment between curves. (c) The scaling of the oscillation frequency $f_{osc}$ for each approach. The filled symbols indicate the fitted region and have a spacing of $1$ dB in power. (d) The simulations' magnon mode amplitudes $c(\kbf)$ for powers 2 dB (bottom, no broadening) and 25 dB (top, with broadening) greater than $P_{osc}$.
}}
    \label{fig:Fig3}
\end{figure}


As the relative power $10\mathrm{\,log_{10}(P_{a}/P_{osc}})$ increases to 14 dB, the oscillations' frequency spectra broaden in the experiment and simulations [Figs. 3(a,b)]. To investigate this, we compare the simulations' magnon mode amplitudes $c(\kbf)$ \cite{qu2020nonlinear} at the relative powers of 2 dB (no broadening) and 25 dB (pronounced broadening) [Fig. 3(d)]. At 2 dB, only the $f_0/2$ modes with the largest coupling strengths $\zeta_{\kbf_i}$ are excited. Note that the coupling is strongest for the modes with wavevectors most misaligned with the static magnetization $\mathbf{M}$. At 25 dB, the weaker-coupled $f_0/2$ modes are also excited, with some even exceeding the amplitude of the strongest-coupled modes. The excited modes also exhibit a wider frequency distribution about $f_0/2$, which may generate the observed broadening. The most straightforward explanation for this transition is the onset of four-magnon scattering of $f_0/2$, opposite-wavevector pairs of magnons at the strongest-coupled modes to such pairs at weaker-coupled modes, which conserves energy and momentum.

At the highest powers, the splitting becomes pronounced enough to introduce negative values of $c_0$ [Figs. 4(a,b)]. As $c_0$ is phase-sensitive, being obtained by mixing $b_0$ with a reference signal, this corresponds to $b_0$ undergoing a phase shift $\phi_0\sim180^\circ$. Phase shifts for the mode $b_i$ arise when its response is dominated by its scattering term [see Eqs. \eqref{eq:db0_dt},\eqref{eq:dbk_dt}], such that it is strongly nonlinear. At these powers, we also observe pronounced oscillations of $c_0$ after turning off the microwave excitation. Notably, these oscillations' amplitudes greatly exceed the steady-state value of $c_0$ during excitation and they persist for roughly 600 ns. Each curve in Figs. 4(a,b) is normalized to the turn-on peak at the highest power, showing the oscillations at excitation turn-on and turn-off to be comparable in size. To understand these observations, we examine the case of strong nonlinearity in our model by numerically solving Eqs. \eqref{eq:db0_dt},\eqref{eq:dbk_dt}, which provides the evolution of the modes' amplitudes as well as their phases. We calculate the modes' phase shifts $\phi_i(t)$ by using our general definition $b_i(t)=|b_i(t)| e^{i(\omega_i t+\theta_i+\phi_i(t))}$:
\begin{align}
    \phi_i(t)=\frac{1}{i}\mathrm{ln}\left(\frac{b_{i}(t)}{|b_i(t)|}e^{-i(\omega_it+\theta_i)}\right).  \label{eq:phi_i_def} 
\end{align}
The evolution of the amplitudes and phase shifts is shown in Figs. 4(c,d) and Figs. 4(e,f); we utilize $|\phi_i(t)|$ for simplicity. For consistency with the experiment, we plot $c_0$ instead of $|b_0|$, taking $c_0(t)=\mathrm{Re}(b_0(t) e^{-i(\omega_0t+\theta_0)})$.


\begin{figure}[t]
    \includegraphics{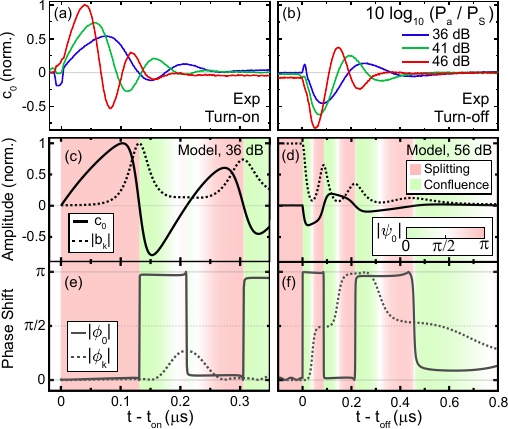}
    \caption{ \selectfont{FIG. 4. (a,b) The experimental results for $c_0(t)$ at high powers when turning on and off the microwave excitation. Each curve is normalized to the peak amplitude at turn-on for the relative power of 46 dB. (c,d) The magnon amplitudes $c_0$, $|b_k|$, with each time color-coded for the corresponding value of $|\psi_0|$. Note that $c_0$ at turn-off is instead normalized to the peak amplitude at turn-on for the relative power of 56 dB. (e,f) The corresponding phase shifts $|\phi_{0,k}|$.
}}
    \label{fig:Fig4}
\end{figure}


After turning on the excitation [Figs. 4(c,e)], we find $\pi$ phase shifts of $\phi_0$ at $c_0=0$, with these phase shifts triggering variation in $\phi_k$. We first examine the influence of the $\pi$ phase shift on the amplitudes' equations of motion. Whereas $c_i(t)$ is the mode's amplitude for $\phi_{0,k}=0,0$, we define $c'_i(t)$ as the mode's amplitude for $\phi'_{0,k}=\pi,0$. Substituting $b_i(t)=c'_i(t) e^{i(\omega_i t+\theta_i+\phi'_i)}$ into Eqs. \eqref{eq:db0_dt},\eqref{eq:dbk_dt} yields the amplitudes' new equations of motion given the $\pi$ phase shift:
\begin{align}
\dot{c}'_0&=-\eta_0 c'_0+N_k \bar{\zeta} c'^2_k-\nu h_{a}, \label{eq:dc0_dt_confluence} \\
\dot{c}'_k&=-\bar{\eta}_k c'_k -\bar{\zeta} c'_k c'_0. \label{eq:dck_dt_confluence}
\end{align}
Comparison with Eqs. \eqref{eq:dc0_dt},\eqref{eq:dck_dt} shows that the $\pi$ phase shift switches the sign of both the microwave field term $\nu h_a$ and the three-magnon scattering terms $ \sim \Bar{\zeta}$. From the latter, it is evident that these phase shifts correspond to reversals in the three-magnon scattering direction between splitting and confluence. The reversals to confluence explain why $|c_0(t)|$ increases with time despite being damped by the now out-of-phase microwave field for $c_0<0$ [Figs. 4(a,c)]. Furthermore, they explain the variation of $\phi_k$: with $|b_k(t)|$ being suppressed by confluence, its three-magnon scattering dominates over its linear terms such that $b_k$ enters the strong nonlinearity regime.

These reversals are more directly evident after turning off the microwave excitation, where $|c_0(t)|$ undergoes a pronounced increase in time despite the absence of the excitation field. Furthermore, without the microwave field to drive $\phi_0$ back to $0^\circ$, the variations of $\phi_k$ evolve into $90^\circ$ phase shifts of $b_k$, generating additional reversals [Figs. 4(d,f)]. The $180^\circ, 90^\circ$ phase shifts of $b_0,b_k$ are those required to reverse the scattering direction, with the factor-of-two difference being due to the same factor difference in their frequencies. These reversals explain both the pronounced oscillations of $c_0$ at turn-off and the oscillations' enhancement with microwave power: the scattering at turn-off, and hence the reversals, are driven by the steady-state values of $c_0,c_k$ during turn-on, where $c_k\propto\sqrt{h_{a}-h_{S}}$ at steady state [see Eq. \eqref{eq:nonlinear_ckf_def} in SM]. The model's turn-off oscillations of $c_0$ are much weaker than those in the experiment, hence the use of the higher relative power of 56 dB. This may be because of our use of a linear damping term $-\eta_i b_i$, which is insensitive to phase shifts, instead of the full Gilbert damping term.

To determine how the scattering direction evolves in time, we consider the relative phase $|\psi_i(t)|$ between $b_i$ and the three-magnon scattering term in its equation of motion. As with the linear damping term $-\eta_i b_i$, the damping of $b_i$ by scattering corresponds to the scattering term being $\pi$ out-of-phase with $b_i$. Conversely, the scattering term drives $b_i$ when it is in-phase with $b_i$. 
As such, $|\psi_0|=0,\pi$ corresponds to confluence and splitting, respectively, and we can calculate $|\psi_0(t)|$ to determine the evolution of the scattering direction. Note that $|\psi_0(t)|$ and $|\psi_k(t)|$ are found to mirror each other about $\pi/2$ as expected, such that one mode is being driven by scattering while the other mode is being damped. From Eq. \eqref{eq:db0_dt}, $\psi_0(t)$ takes the form
\begin{align}
    \psi_0(t)=\frac{1}{i}\mathrm{ln}\left(\frac{b_{0}(t)/|b_0(t)|}{b_k^2(t)/|b_k^2(t)|}\right). \label{eq:psi_i_def}
\end{align}
$|\psi_0(t)|$ corresponds to the color-coding in Figs. 4(c-f). The switching of $|\psi_0(t)|$ between $0,\pi$ aligns with the phase shifts and the transitions between growth and decay of $b_k(t)$ as expected, verifying that the relative phase $|\psi_i(t)|$ indicates the three-magnon scattering direction.

In summary, for three-magnon scattering of ferromagnetic resonance, we observe a regime that hosts transient oscillations of the magnon populations, with the transient behavior being highly dependent on the excitation power. At high excitation powers, we find that the scattering generates significant phase shifts of the magnons and that these phase shifts correspond to reversals between three-magnon splitting and confluence. Such reversals also occur upon turning off the excitation, generating prolonged and coherent oscillations. Our model captures these behaviors. These findings shed light on the transient behavior of this instability, and reveal the nontrivial interplay between three-magnon scattering and the magnons' phases.


The authors thank Aneesh Venugopal for fruitful discussion on efficient computation in the micromagnetic simulations and Cody Schimming for valuable mathematical insight. The Minnesota Supercomputing Institute (MSI) provided resources that contributed to the research results reported within this article. The authors acknowledge support by SMART, a center funded by nCORE, a SRC program sponsored by NIST. The authors also acknowledge support by DARPA under Grant W911NF-17-1-0100, MINT at Minnesota, and the NSF XSEDE through Allocation No. TG-ECS200001. 

\hfill

T. Q. and A. H. contributed equally to this work.



%

\newpage

\large 

\begin{center}
\textbf{Supplementary Material: Oscillations and confluence in three-magnon scattering of ferromagnetic resonance}
\end{center}

\normalsize

\section{Experimental methods}\label{sec:exp_supp}


The experiment was performed on a commercially obtained \cite{matesysite} 3 $\mu$m-thick YIG film grown by liquid phase epitaxy on a GGG substrate. A sample of 2 mm width and 5 mm length was obtained via a wafer saw, then centered (with the YIG side facing down) on top of a homemade microstrip waveguide of 3.4 mm strip width. We utilize a microstrip waveguide of larger width than the magnetic sample for microwave field homogeneity. This suppresses the excitation of millimeter-wavelength magnon modes, yielding an isolated FMR peak with no prominent satellite peaks and a linewidth $\Delta H \approx$ 0.3 Oe. After characterizing the resonance through field-swept measurements, via a lock-in amplifier, time-resolved measurements at zero field and ferromagnetic resonance were performed with an oscilloscope. These measurements utilized homodyning spectroscopy as detailed below.

For our homodyning spectroscopy circuit, a microwave source provides a signal at the desired FMR frequency, which is then split into the RF and LO branches necessary for mixing. The LO branch goes directly to the LO port of the mixer to drive the mixer. Meanwhile, the signal in the RF branch is set to the desired power through the attenuator, and then converted via a switch from a steady signal to an $8\,\mu$s duration, $50\,\%$ duty cycle pulsed signal. The pulsed RF signal is then fed into the microstrip waveguide, which is placed between the poles of an electromagnet. Due to inductive coupling to the waveguide, the magnetic response of the YIG can be measured as a voltage superimposed on the transmitted RF signal. The RF signal is then fed into a phase shifter such that the phases of the RF and LO branches are matched at their inputs to the mixer. Isolators are placed at the RF and LO inputs of the mixer to suppress reflections from the mixer. Due to the phase- and frequency-matching between the RF and LO ports, the output of the mixer corresponds to the envelope of the RF signal. This output is fed into a low-pass filter and terminated at the input of an oscilloscope or lock-in amplifier. To isolate the absorption voltage $V_{abs}(t)$, which corresponds to the FMR amplitude $c_0(t)$, the RF signals' envelopes at FMR are subtracted from those at zero field. 

For the time-resolved measurements, the captured waveform is an average of 512 waveforms. The captured waveform contains several pulses, which are then averaged over as well. Lastly, small moving averages were applied to the data. The moving average window size went incrementally from 5 points at high powers to 50 points to low powers. Given the time resolution of 2-5 ns, these window sizes were selected so as to not impact the time behavior on timescales that are relevant to this study. 

To directly compare the FMR amplitudes between different approaches, we normalize them to the absorptive susceptibility $\chi^{''}(t)=c_0(t)/h_a$, which is normalized in turn to its steady-state value in the linear regime $\chi^{''}_L$. In the experiment, $\chi^{''}(t)$ was obtained by treating the absorption voltage $V_{abs}(t)$ and the applied microwave voltage amplitude $V_{a}$ as proxies for $c_0$ and $h_a$. We also observe a voltage offset $V_{0}$, which likely arises from leakage from the LO port to the RF port. This yields the relation $\chi_{exp}^{''}= V_{abs}/(V_{a}-V_{0})$. 

The experimental value of $\eta_0$ (defined here as $\eta_{0,exp}$) at each FMR frequency was found by fitting the linear behavior $c_0(t)\propto(1-e^{-\eta_0t})$ to the experimental data in the linear regime at multiple powers and averaging the fitted values of $\eta_0$. Via the definition of $\eta_i$ in the main text, the associated Gilbert damping constant is then calculated via the equation $\alpha_{exp}=\eta_{0,exp}/\omega_0 \epsilon_0$. The simulations' timescales were then re-scaled by the value $\alpha_{sim}/\alpha_{exp}$, as an artificially large $\alpha_{sim}$ was used for computational efficiency. In addition, the measured value of $\eta_{0,exp}$ was the value of $\eta_0$ utilized in the numerical solving of our model. We also utilized the associated value of $\bar{\eta}_k$, which was calculated by treating the damping constant as independent of wavevector and taking $\epsilon_k=1$ such that $\bar{\eta}_k=\eta_{0,exp}/2\epsilon_0$.

\section{Derivation of semianalytical model} \label{Derivation}\label{sec:model_supp}

\begin{figure}[h]
    \centering
    \includegraphics{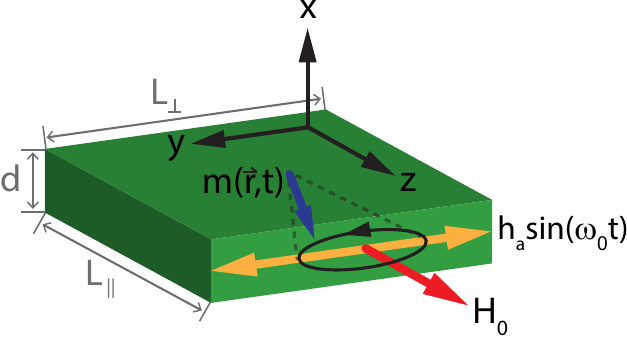}
    \caption{\fontsize{10}{12} \selectfont{Fig. S1. Sample and field geometry. For the perpendicular pumping configuration, the static field $H_0$ is applied in-plane and perpendicular to the microwave field $h_a(t)$. $H_0$ saturates the static magnetization to $M_s$ and sets its direction. $h_a$ dynamically excites the magnetization, causing it to precess around the direction of its static component at an angle proportional to the magnon population. This precession is elliptical in thin films due to the dynamic demagnetization field. }}
    \label{fig:FigS1}
\end{figure}

We start with the Landau-Lifshitz equation
\begin{align}
    \dot{\mathbf{m}}=-\gamma (\mathbf{m}\times \mathbf{H_{eff}}), \label{eq:LLeqn}\\
    \mathbf{H_{eff}}=\mathbf{H_0}+\mathbf{h_a(t)}+\mathbf{h_{ex}}+\mathbf{h_{d}}
\end{align}
where $\mathbf{m}$ is the unit vector of the magnetization, $\gamma$=17.7 MHz/Oe is the gyromagnetic ratio, $\mathbf{H_{eff}}$ is the effective magnetic field, $\mathbf{H_0}$ is the applied static field, $\mathbf{h_a(t)}$ is the applied microwave field, $\mathbf{h_{ex}}$ is the exchange field, and $\mathbf{h_{d}}$ is the demagnetization field. We want to re-express Eq. \eqref{eq:LLeqn} in terms of the circular magnon modes $a_{\kbf},a^*_{-\kbf}$, where \cite{suhl1957theory_supp}
\begin{align}
m^+=m_x+im_y=\sum_\kbf{a_{\kbf}e^{i\kbf \cdot \mathbf{r}}}\delta,\\
m^-=m_x-im_y=\sum_\kbf{a^*_{-\kbf}e^{i\kbf \cdot \mathbf{r}}}\delta, \nonumber \\
\delta=\frac{\Delta k_\parallel \Delta k_\perp}{(2\pi)^2}, \nonumber \\
\Delta k_{\parallel,\perp}=\pi/L_{\parallel,\perp}. \nonumber
\end{align}
The summations are over all available in-plane wavevectors $\kbf=(k_\parallel,k_\perp)$, where $k_\parallel,k_\perp$ are the wavevector components parallel and perpendicular to the static magnetization direction, respectively. $\Delta k_\parallel, \Delta k_\perp$ are the magnon mode spacings parallel and perpendicular to the static magnetization direction. From Eq. \eqref{eq:LLeqn}, defining $\omega_i=\gamma H_{eff,i}$ and $\tilde{D}=\gamma D$  (with $D=0.48 \: \mathrm{Oe}\: \mathrm{\mu m^2}$ being the exchange stiffness) yields
\begin{align}
\mathbf{\dot{m}_x}=m_z\omega_y-m_y\omega_z, \\
\mathbf{\dot{m}_y}=m_x\omega_z-m_z\omega_x, \nonumber
\end{align}
where
\begin{align}
\omega_x=\gamma(h_{ex,x}-h_{d,x})=\tilde{D}\nabla^2m_x-\omega_{d,x}, \\
\omega_y=\gamma(h_{ex,y}-h_{d,y}+h_a(t))=\tilde{D}\nabla^2m_y-\omega_{d,y}+\omega_a(t), \nonumber \\
\omega_z=\gamma(h_{ex,z}-h_{d,z}+H_0)=\tilde{D}\nabla^2m_z-\omega_{d,z}+\omega_H. \nonumber
\end{align}
The demagnetization terms take the form \cite{kalinikos1986theory}, where $d$ is the film thickness,
\begin{align}
\begin{pmatrix}
\omega_{d,x}\\
\omega_{d,y}\\
\omega_{d,z}\\
\end{pmatrix}
=4\pi \gamma M_s \delta \sum_\kbf{
\begin{pmatrix}
(1-G_\kbf) & 0 & 0\\
0 & G_{\kbf} \frac{k_y^2}{k^2} & G_{\kbf} \frac{k_y k_z}{k^2}\\
0 & G_{\kbf} \frac{k_y k_z}{k^2} & G_{\kbf} \frac{k_z^2}{k^2}
\end{pmatrix}
\begin{pmatrix}
m_{\kbf,x}\\
m_{\kbf,y}\\
m_{\kbf,z}\\
\end{pmatrix}
},\\
G_{\kbf}=1-\frac{1-e^{-kd}}{kd}, \nonumber \\
m_{\kbf,x}=\frac{a_\kbf+a^*_{-\kbf}}{2}e^{i\kbf \cdot \mathbf{r}}, \nonumber\\
m_{\kbf,y}=\frac{a_\kbf-a^*_{-\kbf}}{2i}e^{i\kbf \cdot \mathbf{r}}, \nonumber\\
m_{\kbf,z}=0. \nonumber
\end{align}
We define $m_{\kbf,z}=0$ due to only taking the zeroth order term in the expansion $m_z=1-\frac{1}{2}\sum_{\kbf,\kbf'}{a_{\kbf'}a^*_{\kbf'-\kbf}e^{i\kbf \cdot \mathbf{r}}}$, such that $m_z=1$. As such, $m_{\kbf,z}$ corresponds to the Kronecker delta $\delta_{\kbf,0}$. However, the matrix elements vanish at $\kbf=0$ such that the Kronecker delta can be neglected. 

Next, due to its spatial homogeneity, we approximate the microwave field to only couple to the $\kbf=0$ mode and solve for $\dot{m}^+=\dot{m_x}+i\dot{m_y}=\sum_\kbf{\dot{a}_{\kbf}e^{i\kbf \cdot \mathbf{r}}\delta}$. Equating the coefficients of $e^{i\kbf \cdot \mathbf{r}}$ on each side yields the equations of motion for the circularly-precessing magnon modes. We only consider the equations of motion of the circular FMR magnon mode $a_0$ and the half-frequency ($f_0/2$) magnon modes $a_{\kbf_i}$. We have neglected all other modes because, barring higher-order interactions, they are not excited in the first order Suhl instability. The equations of motion of $a_0, a_{\kbf_i}$ are found to be:

\begin{align}
\dot{a}_{\kbf_i} = i(A_{\kbf_i} a_{\kbf_i}+B_{\kbf_i} a^*_{-\kbf_i}) - a_0 \phi_{\kbf_i} (a_{\kbf_i}-a^*_{-\kbf_i}), \label{eq:dak_dt}\\
\dot{a}_0=i(A_0 a_0+B_0 a^*_0) - \sum_{i=1}^{N_k}{a_{\kbf_i} \phi_{-\kbf_i} (a_{-\kbf_i}-a^*_{-\kbf_i})+\omega_a(t)}, \label{da0_dt}\\
A_\kbf=\omega_H+\tilde{D}k^2+2\pi\gamma M_s[1-G_\kbf+G_\kbf \frac{k_y^2}{k^2}], \\
B_\kbf=2\pi\gamma M_s[1-G_\kbf-G_\kbf \frac{k_y^2}{k^2}], \\
\phi_{\kbf}=2\pi\gamma M_s \delta G_{\kbf} \frac{k_{y} k_{z}}{k^{2}}=2\pi\gamma M_s \delta G_{\kbf} \mathrm{sin} \theta_\kbf \mathrm{cos} \theta_\kbf. \label{eq:phi_k}
\end{align}

Note that we only kept terms up to second order in $a$, neglecting higher-order interactions. The $A_\kbf$ and $B_\kbf$ terms correspond to the magnon's angular frequency $\omega_{\kbf_i}$ such that $\omega_{\kbf_i}=\sqrt{A_{\kbf_i}^2-B_{\kbf_i}^2}$. $\phi_\kbf$ is the coupling strength between the circular FMR magnons and the circular magnons at $f_0/2$; the second-order terms, proportional to $\phi_\kbf$, are the three-magnon scattering terms. The summation in Eq. \eqref{da0_dt} captures three-magnon scattering of $a_0$ to the $N_k$ magnon modes in the frequency range $(\omega_0/2-\eta_\kbf)<\omega_{\kbf_i}<(\omega_0/2+\eta_\kbf)$. $\eta_{\kbf}$ is the relaxation rate of the $f_0/2$ modes, whose ellipticity is negligible. The $\omega_a(t)$ term corresponds to the driving from the microwave field.

Next, we perform a Bogoliubov transformation \cite{dobin2003intrinsic_supp} to the magnon eigenmodes; as the dynamic demagnetization field in thin films leads to elliptical magnetization precession, the eigenmodes correspond to the elliptically-precessing magnon modes $b_\kbf, b^*_{-\kbf}$ :
\begin{align}
    \begin{pmatrix}
    b_\kbf\\
    b^*_{-\kbf}
    \end{pmatrix}
    =
    \begin{pmatrix}
    \lambda_\kbf & \mu_\kbf \\
    \mu_\kbf & \lambda_\kbf
    \end{pmatrix}
    \begin{pmatrix}
    a_\kbf \\
    a^*_{-\kbf}
    \end{pmatrix} ;  \\
    \begin{pmatrix}
    a_\kbf\\
    a^*_{-\kbf}
    \end{pmatrix}
    =
    \begin{pmatrix}
    \lambda_\kbf & -\mu_\kbf \nonumber \\
    -\mu_\kbf & \lambda_\kbf
    \end{pmatrix}
    \begin{pmatrix}
    b_\kbf \\
    b^*_{-\kbf}
    \end{pmatrix},\\
    \lambda_\kbf=\mathrm{cosh}(\xi_\kbf), \nonumber\\
    \mu_\kbf= -\mathrm{sinh}(\xi_\kbf), \nonumber \\
    \mathrm{tanh}(2\xi_\kbf)=B_\kbf/A_\kbf. \nonumber 
\end{align}
To perform the Bogoliubov transformation of a given $f_0/2$ mode $a_{\kbf_i}$, we express its equations of motion in matrix form as 
\begin{align}
    \begin{pmatrix}
    \dot{a}_{\kbf_i}\\
    \dot{a}^*_{-{\kbf_i}}
    \end{pmatrix}
    =
    (iL_{0,{\kbf_i}}+\phi_{\kbf_i} L_{1})
    \begin{pmatrix}
    a_{\kbf_i}\\
    a^*_{-{\kbf_i}}
    \end{pmatrix}, \label{eq:ak_matrix}\\
    L_{0,{\kbf_i}}=
    \begin{pmatrix}
    A_{\kbf_i} & B_{\kbf_i}\\
    -B_{\kbf_i} & -A_{\kbf_i}\\
    \end{pmatrix},\\
    L_{1}=
    \begin{pmatrix}
    -a_0 & a_0\\
    a^*_0 & -a^*_0\\
    \end{pmatrix}.
\end{align}
Performing the Bogoliubov transformation of Eq. \eqref{eq:ak_matrix} yields
\begin{align}
    \begin{pmatrix}
    \dot{b}_{\kbf_i}\\
    \dot{b}^*_{-{\kbf_i}}
    \end{pmatrix}
    =\left[i
    \begin{pmatrix}
    \omega_{\kbf_i} & 0\\
    0 & -\omega_{\kbf_i}
    \end{pmatrix}
    +\phi_{\kbf_i} (\lambda_{\kbf_i}+\mu_{\kbf_i})
    \begin{pmatrix}
    \mu_{\kbf_i} a^*_0-a_0 \lambda_{\kbf_i} & \lambda_{\kbf_i} a_0 - \mu_{\kbf_i} a^*_0\\
    \lambda_{\kbf_i} a^*_0-\mu_{\kbf_i} a_0 & \mu_{\kbf_i} a_0 - \lambda_{\kbf_i} a^*_0\\
    \end{pmatrix}\right]
    \begin{pmatrix}
    b_{\kbf_i} \\
    b^*_{-{\kbf_i}}
    \end{pmatrix}.
\end{align}
In the second matrix on the right-hand side, we substitute $a_0, a^*_0$ in terms of $b_0,b^*_0$. Given their frequencies $b_0\propto e^{i\omega_0t}, b^*_0\propto e^{-i\omega_0 t}$, we only retain the terms with the same frequency as that on the left-hand side $b_{\kbf_i} \propto e^{i\omega_0t/2}, b^{*}_{-{\kbf_i}}\propto e^{-i\omega_0t/2}$. Doing so, we obtain
\begin{align}
    \dot{b}_{\kbf_i}=i\omega_{k}b_{\kbf_i}+\zeta_{\kbf_i} b_0 b^*_{-{\kbf_i}},\\
    \zeta_{\kbf_i}=(\lambda_{\kbf_i}+\mu_{\kbf_i})(\lambda_{\kbf_i} \lambda_0+\mu_{\kbf_i} \mu_0)\phi_{\kbf_i}.
\end{align}
$\zeta_{\kbf_i}$ is the coupling strength between the elliptical FMR mode $b_0$ and the given elliptical $f_0/2$ mode $b_{\kbf_i}$. Note that, as an approximation, we have set the angular frequencies $\omega_{\kbf_i}$ of all magnon modes in the frequency range $(\omega_0/2-\eta_\kbf)<\omega_{\kbf_i}<(\omega_0/2+\eta_\kbf)$ to be equal to $\omega_{k}=\omega_0/2$. This approximation is reasonable for low-damping  materials such as YIG, where $\eta_\kbf\ll\omega_0/2$. Repeating these steps for the Bogoliubov transformation for $\dot{a}_0$ yields
\begin{align}
    \dot{b}_0=i\omega_0 b_0-\sum_{i=1}^{N_\kbf} \zeta_{\kbf_i} b^2_{\kbf_i}+(\lambda_0+\mu_0) \omega_a(t),
\end{align}
where the summation is over all $N_k$ elliptical magnon modes $b_{\kbf_i}$ in the frequency range $(\omega_0/2-\eta_\kbf)<\omega_{\kbf_i}<(\omega_0/2+\eta_\kbf)$. 

Next, we account for linear magnon damping by adding the term $-\eta_{\kbf_i}b_{\kbf_i}$, where 
\begin{align}
    \eta_{\kbf_i}=\omega_{\kbf_i} \alpha \epsilon_{\kbf_i}=\omega_{\kbf_i} \alpha\, \frac{1}{\gamma}\left(\frac{\partial \omega}{\partial H}\right)\bigg|_{\kbf_i,H_0}.
\end{align}
$\eta_{\kbf_i}$ corresponds to the relaxation rate of magnon mode $b_{\kbf_i}$, $\alpha$ is the measured Gilbert damping constant, and $\epsilon_{\kbf_i}$ is the ellipticity factor of the magnon mode $b_{\kbf_i}$. 

We take the driving term to be sinusoidal, substituting $\omega_a=\frac{\gamma h_a}{2} e^{-i\pi/2}(e^{i\omega_0 t}+e^{-i\omega_0 t})$ into the equation of motion. We only keep its term proportional to $e^{i\omega_0 t}$ because that is the term that matches the FMR frequency $\omega_0$. This yields the elliptical magnon modes' equations of motion:
\begin{align}
    \dot{b}_0=(i\omega_0-\eta_0)b_0-\sum_{i=1}^{N_k} \zeta_{\kbf_i} b^2_{\kbf_i}+\nu h_a e^{i(\omega_0 t-\pi/2)} \label{eq:db0_dt_supp_Nmode},\\
    \dot{b}_{\kbf_i}=(i\omega_{k}-\eta_{\kbf_i})b_{\kbf_i}+\zeta_{\kbf_i} b_0 b^*_{-\kbf_i}  \label{eq:dbk_dt_supp_Nmode},\\
    \nu=\gamma (\lambda_0+\mu_0)/2.
\end{align}
$\nu$ is the coupling of $b_0$ to the microwave field amplitude $h_a$. This system of $(N_k+1)$ equations can be simplified by taking $\zeta_{\kbf_i}=\Bar{\zeta}$ and $\eta_{\kbf_i}=\Bar{\eta}_k$, their average values over the $N_k$ magnon modes. We also set $b^*_{-\kbf_i}=b^*_{\kbf_i}$, as the splitting and confluence processes affect each mode equally. This causes the equation of motion for each mode $b_{\kbf_i}$ to be equivalent. We also take each mode's initial value, its thermal amplitude, to be equivalent (which is a good approximation, as $\eta_{\kbf_i} \ll \omega_{\kbf,i}$). Making each $b_{\kbf_i}$ mode equivalent in this way reduces their $N_k$ equations of motion to a single equation of motion of an effective $f_0/2$ mode $b_k$. This simplifies our model to the two equations of motion:
\begin{align}
    \dot{b}_0=(i\omega_0-\eta_0)b_0-N_k \Bar{\zeta} b^2_k+\nu h_ae^{i(\omega_0 t-\pi/2)}, \label{eq: db0_dt_supp_1mode}\\
    \dot{b}_k=(i\omega_k-\Bar{\eta}_{k}) b_k +\Bar{\zeta} b^*_{k} b_0. \label{eq: dbk_dt_supp_1mode}
\end{align}
$b_0$ is the response of the elliptical FMR mode and $b_k$ is the response of the effective $f_0/2$ mode. Note that these are the same equations as those stated in Eqs. (1,2) of the main text.

\section{Derivation of thermal magnon amplitude}\label{sec:therm_supp}

To numerically solve our model we must calculate the magnon mode's initial values, which correspond to their thermal amplitudes. This is done by bridging our semiclassical magnon modes $b_{\kbf,c}, b^*_{-\kbf,c}$ with their quantum-mechanical counterparts $b_{\kbf,q}, b^\dagger_{-\kbf,q}$ \cite{dobin2003intrinsic_supp}, whose thermal amplitudes are given by the Bose-Einstein distribution function. Ref. \cite{dobin2003intrinsic_supp}, for small magnon mode amplitudes, provides the approximation 
\begin{align}
    M^+\approx B\sum_\kbf(\lambda_\kbf b_{\kbf,q}+\mu_\kbf b^{\dagger}_{-\kbf,q})e^{i\kbf \cdot \mathbf{r} }\delta;\\
    B=\sqrt{2\hbar\gamma M_s/V},\\
    \delta=\frac{\Delta k_\parallel \Delta k_\perp}{(2\pi)^2}.
\end{align}
$V=L_\parallel L_\perp d$ is the sample size in units of $\mathrm{cm}^3$, and $\Delta k_\parallel, \Delta k_\perp=\pi/L_{\parallel,\perp}$ are the magnon mode spacings parallel and perpendicular to the magnetization, respectively. For our semiclassical formulation,
\begin{align}
    M^+=M_s m^+=M_s \sum_\kbf(\lambda_\kbf b_{\kbf,c}+\mu_\kbf b^{*}_{-\kbf,c})e^{i\kbf \cdot \mathbf{r} }\delta.
\end{align}
Equating the two forms of $M^+$, we find that $b_{\kbf,c}=\frac{B}{M_s} b_{\kbf,q}$, yielding the thermal amplitude of the semiclassical magnon modes:
\begin{align}
    \left(b_{\kbf,c}\right)_{th}=\frac{B}{M_s}\left(b_{\kbf,q}\right)_{th}=\sqrt{\frac{2\hbar\gamma f_{BE}(\kbf)}{M_s V}},\\
    f_{BE}(\kbf)=(e^{hf_\kbf/k_BT}-1)^{-1}.
\end{align}
Note that we take $\left(b_{\kbf,c}\right)_{th}=\left(c_{\kbf,c}\right)_{th}$. 

\section{Computational methods}\label{sec:methods_supp}

\subsection{Numerical solutions of model}

For the numerical solutions to our model, the relevant magnon modes are found through calculating the dispersion only for positive ($k_\parallel$, $k_\perp$), where $\parallel$ and $\perp$ correspond to the magnon wavevector component parallel and perpendicular to the static magnetization direction. This prevents double-counting with the modes at negative ($k_\parallel$, $k_\perp$), and the modes in the other quadrants are negligible due to having a negative coupling strength [as seen through the coupling strength being proportional to $\mathrm{sin}(\theta_{\kbf_i})\mathrm{cos}(\theta_{\kbf_i})$ in Eq. \eqref{eq:phi_k} and evidenced in the simulation results of Ref. \cite{Qu2020}]. 

For the determination of the relevant magnon modes and the numerical solving of their equations of motion, we utilize the value of $\eta_0$ measured in the experiment, which we define as $\eta_{0,exp}$. We measure $c_0(t)$ at several powers in the linear regime and fit the time-dependence to the standard form $c_0(t)\propto(1-e^{-\eta_0t})$, yielding the fitted value of $\eta_0$ for each power. $\eta_{0,exp}$ corresponds to the average of these fitted values of $\eta_0$. From the definition $\eta_i=\omega_i \alpha \epsilon_i$, we can calculate the associated Gilbert damping constant $\alpha_{exp}$ for a given FMR frequency $f_0$ and the associated ellipticity factor $\epsilon_0$. $\alpha_{exp}$ is the value used to calculate the associated values of $\eta_{\kbf_i}$ to obtain $\bar{\eta}_k$.

We take the $f_0/2$ magnon modes in our calculated dispersion to be those in the frequency range $\omega_0/2-\eta_k<\omega<\omega_0/2+\eta_k$, where we approximate negligible ellipticity such that $\eta_k=\omega_0 \alpha_{exp} /2$. The number of modes in this range is our calculated value of $N_k$, and we calculate their individual coupling strengths and their relaxation rates $\zeta_{\kbf_i},\eta_{\kbf_i}$ to obtain the average values over these modes $\bar{\zeta},\bar{\eta}_k$.  

For numerically solving our model, we need to utilize values of $h_a$ such that the relative excitation $h_a/h_S$ is consistent with that in the experiment and simulations; $h_S$ is the threshold value of $h_a$ for the instability. Our linearization of the equations of motion of $c_0,c_k$ provides the analytical value $h_S=\eta_0 \bar{\eta}_k/\nu \bar{\zeta}$ (see Sec. \ref{sec:linearize_supp}), which we find to be valid for the numerical solving of $c_0,c_k$. However, for numerically solving $b_0,b_k$, this value of $h_S$ is found to no longer be valid, potentially due to the influence of the phase degree of freedom on the onset of the instability. By numerically solving for $b_0,b_k$ with trial values of $h_a$ to find $h_S$, we find that the associated value of $h_S$ increases by a factor of approximately 5.6, or 15 dB in power. 

Lastly, to numerically solve our model, we need to calculate the initial values of $b_i,c_i$. We take these to correspond to their thermal amplitudes, which are calculated through the Bose-Einstein distribution. To do so, we connect our semiclassical formulation with the quantum-mechanical formulation in Ref. \cite{dobin2003intrinsic_supp} (see Sec. \ref{sec:therm_supp}). With these calculated values for the parameters and initial values, we numerically solved the equations of motion through the ode45 function in MATLAB. For the derivation of our model, see Sec. \ref{sec:model_supp}.

\subsection{Micromagnetic simulations}

Micromagnetic simulations were performed using the Landau-Lifshitz-Gilbert (LLG) equation \cite{tang2010influence,natekar2017calculated}
\begin{align}
  \mathbf{\dot{m}}=-\gamma\mathbf{m}\times(\mathbf{H}_{\mbox{\footnotesize eff}}+\boldsymbol{\xi}(T))+\alpha\mathbf{m}\times\mathbf{\dot{m}} \,,\\
  \mathbf{H_{eff}}=\mathbf{H_0}+\mathbf{h_a(t)}+\mathbf{h_{ex}}+\mathbf{h_{d}},
\end{align}
where $\mathbf{m}(\mathbf{r},t)$ is the unit vector of the local magnetization and $\gamma=17.7$ MHz/Oe is the gyromagnetic ratio. $\mathbf{H}_{\mbox{\footnotesize eff}}$ is the effective field including the in-plane static field $\mathbf{H_0}$, the applied microwave field $\mathbf{h_a}(t)$, the exchange field $\mathbf{h_{ex}}$, and the demagnetization field $\mathbf{h_d}$. $\boldsymbol{\xi}(T)$ is the thermal fluctuation field and $\alpha$ is the Gilbert damping constant. $\boldsymbol{\xi}(T)$ is dependent on the temperature $T$ \cite{Brown1963,Liu2017} and corresponds to a white-noise field whose amplitude $\xi$ is determined by the equation 
\begin{equation}\label{eq:power_field_conversion}
\xi(T)=\sqrt{\frac{2k_BT\alpha}{\gamma V dt M_s}} \,.
\end{equation}
$k_B$ is the Boltzmann constant, $V$ is the cell volume in the simulation, $dt$ is the utilized time increment, and $M_s$ is the saturation magnetization.

The Cartesian coordinate system and magnetic field orientations are the same as those shown in Fig. S1 in Sec. \ref{sec:model_supp}. As the microwave pumping frequency $f_0$ is $\sim$ 1 GHz, the wavelength of the microwaves is $\sim$ 1 cm. Compared to our simulated system of $\sim 1$ mm lateral size, the microwave field can be treated as a uniform ac magnetic field oscillating at the frequency $f_0$. 

For our sample material, yttrium iron garnet, we use the following parameters for the simulation: the gyromagnetic ratio $\gamma$=17.7 MHz$/\mathrm{Oe}$, the exchange constant A=3.5$\times$10$^{-7}$ erg/cm, and the saturation magnetization $M_s$=130 emu/cm$^3$. Note that, to improve computational efficiency via reduction of the instability's timescales, we utilized an inflated value of the Gilbert damping constant for the simulations, setting $\alpha_{sim}=0.005$. This also inflates the corresponding value of $h_{osc}$, necessitating the normalization to $h_{osc}$ in our definition $\tilde{h}=(h_a-h_{osc})/h_{osc}$ in the main text.

A custom parallel-computing code based on CUDA \cite{Aneesh2019} is used to optimize the computational performance. The lateral size of the system under study is 30 $\mu$m $\times$ 30 $\mu$m, and the thickness is 3 $\mu$m. This system is discretized into cells with the dimensions 100 nm $\times$ 100 nm $\times$ 3 $\mu$m. For each timestep $dt=50$ fs, the simulation numerically solves the LLG equation for each cell to compute its magnetization. From this time- and space-resolved data, we can calculate the time-resolved magnon mode amplitudes $c(\kbf,t)$, as done in ref. \cite{Qu2020}. 

As the timescales of the instability are inversely proportional to $\alpha$, the inflation of $\alpha$ necessitates re-scaling of the times in the simulation. As such, to enable direct comparison of our simulations with both our experiment and the numerical solutions to our model, we re-scale its time values by the factor $\alpha_{sim}/\alpha_{exp}$, where $\alpha_{exp}$ is calculated as described in the previous subsection.

\section{Data at 2.5 GHz FMR frequency}\label{sec:2_5GHz_Supp}

In Fig. S2, the top row ($f_0=1.5$ GHz) is the same data as that in Fig. 2 of the main text, with the corresponding data at $f_0=2.5$ GHz shown in the bottom row. For a given FMR frequency, each approach shows strong qualitative agreement for a given relative power $P_a/P_S$. An inset of the highest-power experimental data at each frequency is provided to show the crossings of $c_0$ through zero, corresponding to $180^\circ$ phase shifts of $b_0$.

\begin{figure}[h]
    \centering
    \includegraphics[width=\textwidth]{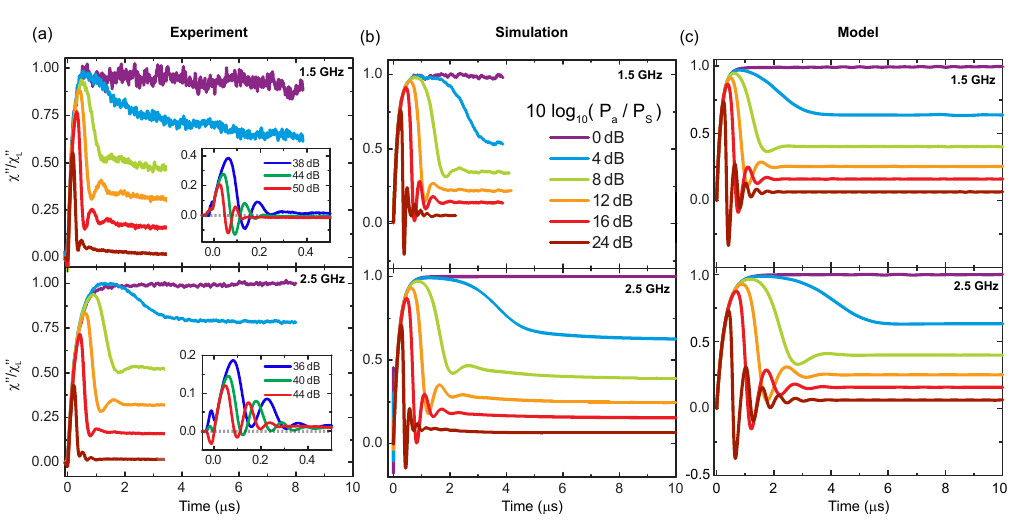}
    \caption{\fontsize{10}{12} \selectfont{FIG. S2 Full comparison of the instability's nonequilibrium behavior from (a) the experiment, (b) the simulations, and (c) the numerical solutions to Eqs. (3,4). The top row ($f_0=1.5$ GHz) is the same data as in Fig. 2 of the main text. The corresponding data at $f_0=2.5$ GHz is shown in the bottom row. }}
    \label{fig:2_5Qual}
\end{figure}

\newpage

\begin{figure}[h]
    \centering
    \includegraphics[width=\textwidth]{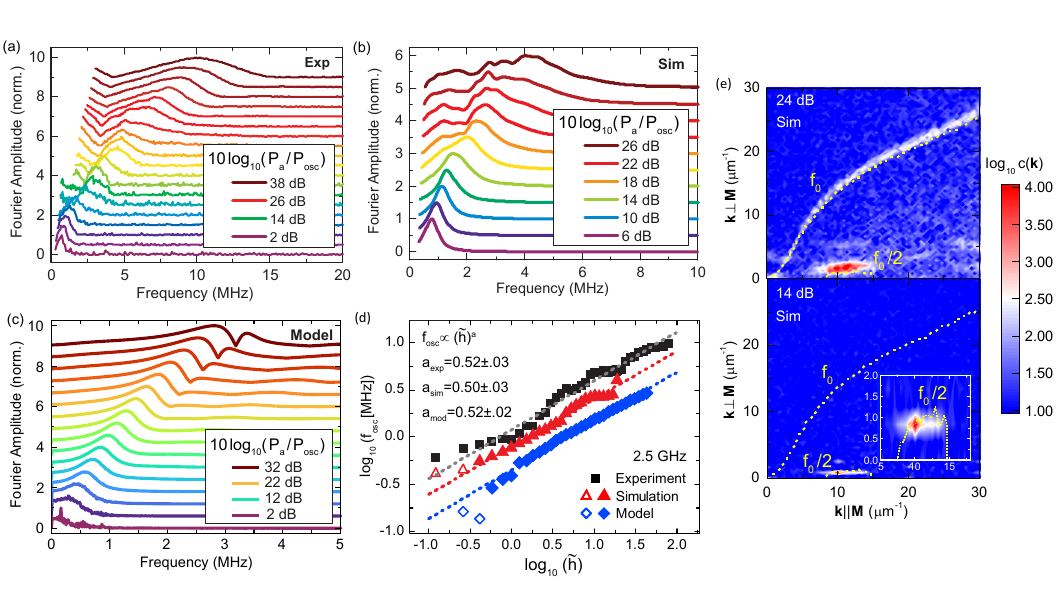}
    \caption{\fontsize{10}{12} \selectfont{FIG. S3 Oscillation analysis for $f_0=$2.5 GHz. (a-c) Normalized and offset frequency spectra of the oscillations in (a) the experiment, (b) the micromagnetic simulations, and (c) the numerical solutions to Eqs. (3,4) in the main text. (d) The scaling of the oscillation frequency $f_{osc}$ for each approach. The filled symbols indicate the fitted region, where quantitative agreement of the scaling with our linearized model is observed. (e) The simulated steady-state magnon amplitudes $c(\kbf)$ for powers 14 dB (bottom) and 24 dB (top) above the oscillation threshold.}}
    \label{fig:Osc_anal_2_5}
\end{figure}

Figure S3 is the same layout as Fig. 3 in the main text, but with the data being at $f_0=2.5$ GHz and the inclusion of the oscillations' frequency spectra from the model's numerical solutions. In these numerical solutions, the lopsidedness of the transient behavior at high powers [see Fig. S2(c)] leads to a distortion in their frequency spectra [Fig. S3(c)]. As such, we isolate the analysis of the numerical solutions' frequency scaling to powers below which this distortion dominates the spectra. As at $f_0=1.5$ GHz, we observe quantitative agreement of the scaling of $f_{osc}$ with that predicted by our linearized model over several orders of magnitude in microwave power. Note that, in Fig. S3(d), there is a $1$ dB increment in power between points.

In the experiment and simulations, as observed for $f_0=1.5$ GHz, the oscillations' frequency spectra broadens for relative powers above approximately 14 dB [Figs. S3(a,b)]. Fig. S3(e) shows, for the relative powers of 14 dB (minimal broadening) and 24 dB (pronounced broadening), the steady-state magnon amplitudes $c(\kbf)$ from simulations. As observed at $f_0=1.5$ GHz in the main text, this broadening corresponds to a transition from isolated excitation of the strongest-coupled $f_0/2$ modes to excitation of many weaker-coupled $f_0/2$ modes. The excited modes at $25$ dB exhibit a wider frequency distribution around $f_0/2$, which may cause the observed broadening. Recall that this transition is hypothesized to be four-magnon scattering from strongly- to weakly-coupled $f_0/2$ magnon pairs, conserving energy and momentum.

However, for $f_0=2.5$ GHz, there is additional magnon excitation along the $f_0$ contour (top panel). Similarly, the most straightforward mechanism for this excitation is four-magnon scattering of two FMR magnons to two $f_0$ magnons of equal and opposite wavevectors, which conserves energy and momentum. This four-magnon scattering process corresponds to the second order Suhl instability, which is the relevant instability when magnon modes at $f_0/2$ are unavailable.

\section{Derivation of different regimes and the oscillation scaling}\label{sec:linearize_supp}

In order to analyze the oscillatory regime, we linearize Eqs. (3,4) in the main text. These equations are the elliptical magnon amplitudes' equations of motion for the weakly nonlinear case, where phase shifts are negligible:
\begin{align}
    \dot{c}_0=-\eta_0c_0-N_k\Bar{\zeta} c^2_k+\nu h_a, \label{eq: dc0_dt_supp_1mode}\\
    \dot{c}_k=-\Bar{\eta}_k c_k +\Bar{\zeta} c_k c_0. \label{eq: dck_dt_supp_1mode}
\end{align}
We linearize these equations by examining around their fixed points, which are the values $(c_{0,f},c_{k,f})$ such that $\dot{c}_0=\dot{c}_k=0$. This allows us to treat the second-order time-varying terms as negligible, converting the equations of motion into a linear form.

\subsection{Linearization of the equations of motion}

To examine the equations of motion near a fixed point, we define
\begin{equation}
\begin{pmatrix}
c_0(t)\\
c_k(t)
\end{pmatrix}
=\begin{pmatrix}
c_{0,f}\\
c_{k,f}
\end{pmatrix}
+\begin{pmatrix}
U(t)\\
V(t)\\
\end{pmatrix}, \,\, U(t),V(t)\ll 1.
\end{equation}
Given these definitions, Taylor expanding $\dot{c}_0,\dot{c}_k$ about $(c_{0,f},c_{k,f})$ with the Jacobian
\begin{equation}
J\rvert_{c_{0,f},c_{k,f}}=\left. \begin{pmatrix}
\partial \dot{c}_0/ \partial c_0 & \partial \dot{c}_0/ \partial c_k \\
\partial \dot{c}_k/ \partial c_0 & \partial \dot{c}_k/\partial c_k
\end{pmatrix}\right|_{c_{0,f},c_{k,f}}
=\begin{pmatrix}
-\eta_0  && -2N_k \Bar{\zeta} c_{k,f}\\
\Bar{\zeta} c_{k,f} &&-\Bar{\eta}_k+\Bar{\zeta} c_{0,f}
\end{pmatrix}
\end{equation}
yields
\begin{align}
    \begin{pmatrix}
    \dot{c}_0\\
    \dot{c}_k
    \end{pmatrix}
    =
    \begin{pmatrix}
    \dot{U}\\
    \dot{V}
    \end{pmatrix}
    \approx
    \begin{pmatrix}
    -\eta_0  && -2N_k\Bar{\zeta} c_{k,f}\\
    \Bar{\zeta} c_{k,f} &&-\Bar{\eta}_k+\Bar{\zeta} c_{0,f}
    \end{pmatrix}
    \begin{pmatrix}
    U\\
    V
    \end{pmatrix}.
\end{align}
We then impose the form
\begin{equation}
\begin{pmatrix}
U\\
V
\end{pmatrix}
=
\begin{pmatrix}
a e^{\lambda t}\\
b e^{\lambda t}
\end{pmatrix}
\end{equation}
to generate the eigenvalue equation
\begin{align}
\lambda
    \begin{pmatrix}
    U\\
    V
    \end{pmatrix}
    =
    \begin{pmatrix}
    -\eta_0  && -2N_k\Bar{\zeta} c_{k,f}\\
    \Bar{\zeta} c_{k,f} &&-\Bar{\eta}_k+\Bar{\zeta} c_{0,f}
    \end{pmatrix}
    \begin{pmatrix}
    U\\
    V
    \end{pmatrix},
\end{align}
with solutions of $\lambda$ such that
\begin{equation}
\det(J\rvert_{c_{0,f},c_{k,f}}-\lambda I)=0.
\end{equation}
Solving this equation provides the following solution for the two possible values of $\lambda$:
\begin{equation}
\lambda_{\pm}=\frac{-\rho\pm\sqrt{\rho^2-4\sigma}}{2} \label{eq:lambda_pm}
\end{equation}
\begin{equation}
\rho=\eta_0+\Bar{\eta}_k-\Bar{\zeta} c_{0,f}
\end{equation}
\begin{equation}
\sigma=\eta_0 \Bar{\eta}_k+2N_k(\Bar{\zeta} c_{k,f})^2-\eta_0 \Bar{\zeta} c_{0,f}.
\end{equation}
The above is still general, with insertion of the fixed point values $(c_{0,f},c_{k,f})$ for a given regime providing its associated behavior.

\subsection{Linear regime}
The fixed point for the linear regime corresponds to
\begin{equation}
(c_{0,f},c_{k,f})=(\nu h_a/\eta_0,0).
\end{equation} 
Near this fixed point, one can approximate $c_k^2\approx 0$ such that 
\begin{equation}
\dot{c_0}= -\eta_0 c_0 + \nu h_a.
\end{equation}
Including a constant in the solution of the above equation's homogeneous form and applying the boundary conditions $c_0(0)=0,c_0(\infty)=\nu h_a/\eta_0$ yields the analytical solution for $c_0(t)$, which has the standard time-dependence in the linear regime:
\begin{equation}
c_0(t)=\frac{\nu h_a}{\eta_0}(1-e^{-\eta_0 t}).
\end{equation}
We find, upon fitting this equation's time-dependence to the linear FMR behavior in simulations, that the fitted value of $\eta_0$ closely aligns with its expected value $\eta_0=\omega_0 \alpha_{sim} \epsilon_0$. $\omega_0$ is the FMR angular frequency, $\alpha_{sim}$ is the Gilbert damping constant used in the simulation, and $\epsilon_0$ is the calculated ellipticity factor of the FMR mode. Specifically, the discrepancy between the fitted and defined value of $\eta_0$ is $<0.5\%$ at 2.5 GHz, and is $<5\%$ at 1.5 GHz. The above time-dependence for $c_0(t)$ is that used in fitting the experimental data in the linear regime, yielding the experimental value of $\eta_0$ used in the numerical solving of our model.

Furthermore, inserting the linear regime's fixed point into Eq. \eqref{eq:lambda_pm} yields the Suhl instability's threshold microwave field amplitude $h_S$, the value of $h_a$ such that $\lambda_+$ becomes positive:
\begin{equation}
h_{S}=\eta_0 \Bar{\eta}_k/\nu \Bar{\zeta}. \label{eq: singlemode_hinst}
\end{equation}

\subsection{Nonlinear regime}
The nonlinear regime's fixed point, for which $c_{k,f}$ has a finite value, corresponds to
\begin{equation}
(c_{0,f},c_{k,f})=\left(\frac{\Bar{\eta}_k}{\Bar{\zeta}},\sqrt{\frac{\nu h_a-\eta_0\Bar{\eta}_k/\Bar{\zeta}}{N_k \Bar{\zeta}}}\right). \label{eq:nonlinear_fixedpoints}
\end{equation}
Through the definition of $h_S$, Eq. \eqref{eq:nonlinear_fixedpoints} can be re-expressed in terms of $h_S$. This yields the power-dependence of the steady-state value of $c_k$, $c_k \propto \sqrt{h_a-h_{S}}$, stated in the main text:
\begin{equation}
c_{k,f}=\sqrt{\frac{\nu}{N_k \Bar{\zeta}}(h_a-h_{S})}. \label{eq:nonlinear_ckf_def}
\end{equation}
Solving for $\lambda_\pm$ with the fixed point values in the nonlinear regime yields the eigenvalues
\begin{equation}
\lambda_\pm=-\frac{\eta_0}{2}\pm \sqrt{\eta_0^2/4-2\Bar{\zeta}\nu(h_a-h_{S})}. \label{eq:lambda_pm_nl}
\end{equation}
Note that, as $h_a$ is increased past $h_S$, $\lambda_+$ becomes negative such that the nonlinear fixed point becomes stable. Simultaneously, the linear regime's fixed point becomes unstable. This corresponds to the linear and nonlinear regimes' fixed points exchanging stability as $h_a$ is swept through $h_S$, which is defined as a transcritical bifurcation.

As $h_a$ is increased further, past the critical value $h_{osc}$, it causes the radicand in Eq. \eqref{eq:lambda_pm_nl} to cross through zero and become negative. This corresponds to the nonlinear regime's fixed point transitioning from a stable node ($\lambda_\pm$ being purely real with a negative value) to a stable spiral ($\lambda_\pm$ being complex with a negative real component). This marks the transition to the nonlinear oscillatory regime. Solving for the critical value of $h_a$ that causes the radicand to vanish yields $h_{osc}$:
\begin{equation}
    h_{osc}=h_{S}(1+\frac{\eta_0}{8\Bar{\eta}_k}) \label{eq:h_osc_supp}.
\end{equation}
This is the same form as the definition of $h_{osc}$ stated in the main text. Re-expressing the radicand in terms of $h_{osc}$, we find
\begin{align}
\lambda_\pm=-\eta_0/2\pm i \omega_{osc},\\
\omega_{osc}=2\pi f_{osc}=\frac{\eta_0}{2} \sqrt{2\Bar{\zeta}\nu(h_a-h_{osc})}.
\end{align}

This yields the definition of $f_{osc}$ stated in the main text:
\begin{align}
    f_{osc}=\eta_0 \sqrt{\frac{\Bar{\zeta} \nu}{8\pi^2}(h_a-h_{osc})}.
\end{align}

\section{Analysis of oscillations}\label{sec:osc_analysis_supp}

\begin{figure}[h]
    \centering
    \includegraphics[width=0.75\textwidth]{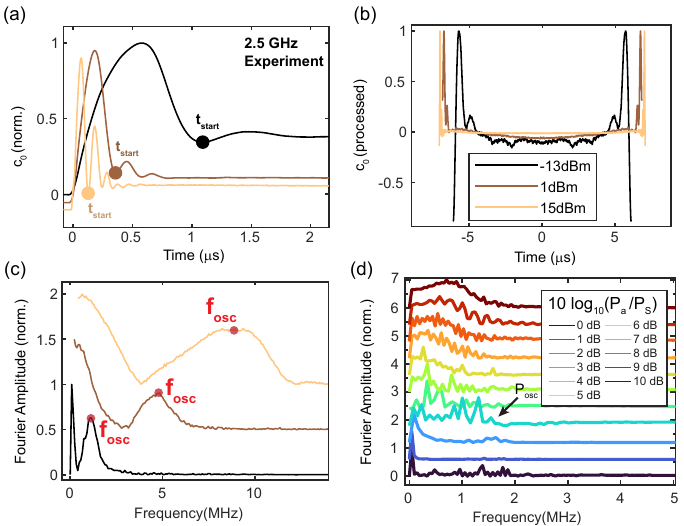}
    \caption{\fontsize{10}{12} \selectfont{FIG. S4. Analysis of the oscillations. (a) Making the window for analyzing the oscillations. (b) Illustration of the processing of the signal. (c) The Fast Fourier Transforms of the processed data in (b) and the determination of their characteristic oscillation frequency $f_{osc}$. (d) Example of the determination of $P_{osc}$ from these spectra.}}
    \label{fig:FigS2}
\end{figure}

The frequency spectra of the oscillations were obtained by doing a fast Fourier transform (FFT) of a window of data of $c_0(t)$. The utilized windows cover the time interval from $t_{start}$, just before the oscillations, to the end of the pulse [Fig. S4(a)]. Before performing the FFT, the window of data was symmetrized and had its average value subtracted to suppress DC contributions to the spectra [Fig. S4(b)]. In addition, for the FFT of the processed data window [Fig. S4(c)], an exclusion of the low-frequency Fourier amplitudes is introduced to better isolate the frequency spectra of the oscillations from DC artifacts. The window of this exclusion is incrementally increased at each power to offset the associated growth of the DC artifact. The oscillation frequency $f_{osc}$ for a given power is defined as that of the largest Fourier amplitude and/or the visible center of the distribution [Fig. S4(c)]. The latter convention is prioritized in the few cases where the former convention provides visibly erroneous results for the characteristic oscillation frequency. 

From the oscillations' spectra, we estimate $P_{osc}$ for each approach as the highest power for which no oscillation-induced structure in the spectra is observed. An example of this determination is shown in Fig. S4(d). The arisal of the oscillations corresponds to the arisal of the broad structure localized at low frequencies, as seen in the 4 dB data at $\approx 0.5$ MHz. 

For calculating the scaling of the oscillation frequency in the experiment, we need to convert both $P_{osc}$ to $h_{osc}$ and $P_a$ to $h_a$ in order to calculate the associated values of $\tilde{h}=(h_a-h_{osc})/h_{osc}$. The conversion factor $\beta$ can be considered as $h=\beta \sqrt{P}$. However, given the normalized form of $\tilde{h}$, $\beta$ will cancel between the numerator and denominator. This allows us to directly solve for $\tilde{h}$ given the absolute powers of $P_a$ and $P_{osc}$: 
\begin{equation}
    \tilde{h}=\frac{\sqrt{P_a}-\sqrt{P_{osc}}}{\sqrt{P_{osc}}}.
\end{equation}

%





\end{document}